\documentclass[seceq]{ptptex}
\newcommand{\meanvalue}[1]{\mbox{$\langle #1 \rangle$}}

\usepackage{graphicx}

\notypesetlogo                       

\title{
Anomalous Heat Conduction in Quasi-One-Dimensional Gases%
}

\author{
Taka H. \textsc{Nishino}%
}

\inst{
Yukawa Institute for Theoretical Physics,
Kyoto University, \\
Sakyo-ku, Kyoto 606-8502, Japan
}

\recdate{June 7, 2007}

\abst{
  From three-dimensional linearized hydrodynamic equations,
  it is found that the heat conductivity is proportional to 
  $(L_x/(L_y^2 L_z^2))^{1/3}$, 
  where $L_x$, $L_y$ and $L_z$ are the lengths of the system along the
  $x$, $y$ and $z$ directions, and we consider the case in which 
  $L_x \gg L_y, L_z$.
  The necessary condition for such a size dependence
  is derived as $\phi \equiv L_x/(n^{1/2} L_y^{5/4} L_z^{5/4}) \gg 1$, 
  where $\phi$ is the critical condition parameter
  and $n$ is the number density.
  This size dependence of the heat conductivity has been confirmed by
  molecular dynamics simulation.
}

\begin{document}

\maketitle

\section{Introduction}
Fourier's law of heat conduction is a relation between
the heat flux $J$ and the temperature gradient.
It is given by $J = - \lambda \nabla T$, where $\lambda$ and $T$ are
the heat conductivity and the temperature, respectively.
Although it is widely believed that Fourier's law is realized
in many situations, recent numerical results suggest that
the heat conductivity diverges in low-dimensional systems 
in the thermodynamic limit \cite{lepri2003}, while it is convergent in
three-dimensional (3D) systems \cite{shimada2000,ogushi2005}.

Narayan and Ramaswamy \cite{narayan2002}
have found that the heat conductivity is proportional to $L^{1/3}$, 
with the system size $L$. They derived this result 
from hydrodynamic equations with thermal fluctuations 
in the 1D limit.
This size dependence of the heat conductivity is also applicable to 1D
chains \cite{mai2006}.
These results have been confirmed in numerical simulations
of the Fermi-Pasta-Ulam (FPU) chain \cite{wang2004}, a 1D gas model
\cite{grassberger2002,cipriano2005} and single-walled carbon nanotubes (SWNT)
\cite{murayama2002,zhang2005}.

In real experiments, of course, it is difficult to realize 
actual one- or two-dimensional systems.
However, it is easy to prepare quasi-one-dimensional (Q1D) systems
in which the role of one direction is dominant, and the length of the
system along that direction is much greater than those along the other
directions.
Similarly, we can construct 
quasi-two-dimensional (Q2D) systems, in which two directions are
dominant, the lengths of the system along those directions are much
greater than those along the other direction.
In this paper, we demonstrate that heat conductivities in Q1D
systems diverge as $\lambda \sim (L_x/(L_y^{2} L_z^{2}))^{1/3}$,
where $L_x$ is the length of the system along the $x$-direction,
and we have $L_x \gg L_y, L_z$.
The necessary condition for this anomalous behavior of 
the heat conductivity is derived as
\begin{equation}
 \phi \equiv
  \frac{L_x}{n^{1/2} L_y^{5/4} L_z^{5/4}}
  = \left( \frac{L_x}{\sqrt{L_yL_z}} \right)^{3/2} \frac{1}{\sqrt{N}} 
  \gg 1 ,
\end{equation}
where $n$ is the number density, $N$ is the number of particles in the system,
and $\phi$ is the critical condition parameter.
We also find that Q2D systems diverge as
$\lambda \sim \sqrt{(\ln (L_x^2+L_y^2))/L_z}$, where
$L_x$ and $L_y$
are much larger than $L_z$.
The necessary condition to obtain this behavior is 
$((L_x^2 + L_y^2) / (n L_z^5)) \ln [\sqrt{L_x^2 + L_y^2}/(t_0 c_0) ]
\gg~1$, where $t_0$ is the mean free time and $c_0$ is the velocity of sound.

\section{Derivation of the long-time tail from the hydrodynamic equations}

The divergence of the transport coefficient in the thermodynamic limit
originates from the long-time tail of the time correlation function, 
which has been confirmed by molecular dynamics (MD) simulations 
with hard spheres \cite{alder1970}.
For the theoretical
calculation of the long-time tail near the equilibrium state,
the mode-coupling theory \cite{pomeau1975} is a powerful tool.
Amongst several methods for calculating the long-time tail
in the mode-coupling theory, 
here, we adopt the hydrodynamic approach developed by Ernst
\emph{et al.} \cite{ernst1971,ernst1976}.

From the Green-Kubo formula \cite{kubo1957,zwanzig1965},
the heat conductivity $\lambda$ can be calculated as
\begin{equation}
 \label{tc}
  \lambda = \beta T^{-1} \lim_{t_c \to \infty}
  \int_0^{t_c} dt \,\, C_\lambda  \left(t \right),
\end{equation}
where $C_\lambda (t)$ is the time correlation function of
the heat flux $J$ and $\beta = (k_B T)^{-1}$ with the Boltzmann constant
$k_B$.
Here, we define the $x$-component of the heat flux $J$ as
\begin{equation}
 \label{jj}
 J \equiv \sum_i
 \left[ \left(\frac{m}{2} v_i^2 - \frac{5}{2} k_B T \right) v_{ix}
   + \frac{1}{2} \sum_{j \neq i}
   \left(\Phi(\boldsymbol{r}_{ij}) v_{ix} - r_{ij,x}
    \frac{\partial \Phi(\boldsymbol{r}_{ij})}{\partial \boldsymbol{r}_{ij}}
    \cdot \boldsymbol{v}_i
   \right)
  \right],
\end{equation}
where $v_{ix}$ is the $x$ component of the velocity of the particle $i$
in a fluid of $N$ particles, $m$ is the mass of a particle,
$r_{ij,x}$ is the $x$ component of the displacement vector between
 particles $i$ and $j$, and $\Phi(\boldsymbol{r})$ is the
intermolecular pair potential.
From this point, we consider only the kinetic parts, $J_K$, of $J$, given by
\begin{equation}
 \label{jk}
  J^K \equiv \sum_i
  \left(\frac{m}{2} v_i^2 - \frac{5}{2} k_B T \right) v_{ix} ,
\end{equation}
and the corresponding correlation function.
Through simulations, as reported in~\S\ref{simulation},
we have verified that the error introduced by ignoring the remaining
terms in \eqref{jj} is negligibly small. 

Although the Green-Kubo formula can only be proven for infinite systems,
if we apply it to finite systems, we need to introduce
upper bound of the integral $t_c$, which represents 
the typical transit time, $t_c \sim L/c_0$,
without taking the limit $t_c \to \infty$ \cite{lepri2003}.

To calculate the time correlation function, we need to solve the
$d$-dimensional linearized hydrodynamic equations
\begin{equation}
 \label{lhe}
 \begin{split}
 \frac{\partial n (\boldsymbol{r},t )}{\partial t} = & -n \nabla \cdot \boldsymbol{u}
  (\boldsymbol{r},t),
  \\
  \frac{\partial \boldsymbol{u} (\boldsymbol{r},t )}{\partial t} = &
  \nu \nabla^2 \boldsymbol{u} (\boldsymbol{r},t )
  + \left(D_l -\nu \right) \nabla
  \left(\nabla \cdot \boldsymbol{u} (\boldsymbol{r},t ) \right)
  \\
 &
  - \frac{c_0^2}{\gamma} \left(\frac{\nabla n (\boldsymbol{r},t)}{n}
  + \alpha_p \nabla T (\boldsymbol{r},t) \right) ,
  \\
 \frac{\partial T \left(\boldsymbol{r},t \right)}{\partial t} = &
  -\frac{\gamma - 1}{\alpha_p} \nabla \cdot \boldsymbol{u} \left(\boldsymbol{r},t \right)+\gamma D_T
  \nabla^2 T \left(\boldsymbol{r},t \right).
 \end{split}
\end{equation}
Here, $\boldsymbol{u}$ is the velocity field and
\begin{equation}
 \label{def-variable}
 \begin{split}
 \gamma &= \frac{c_p}{c_V}, \,
  c_0 = \left[\frac{\gamma}{m} \left(\frac{\partial p}{\partial n}\right)_T
  \right]^{1/2}, \,
  \alpha_p = - \frac{1}{n}  \left(\frac{\partial n}{\partial T} \right)_p, \\
 \nu &= \frac{\eta}{nm},\,
  D_T = \frac{\lambda}{n c_p},\,
  D_l = \frac{2  \left(d-1 \right) \eta + d \xi}{d n m}, 
 \end{split}
\end{equation}
where $p$ is the pressure,
and $c_p$ and $c_V$ are
the heat capacities per particle at constant
pressure and at constant volume, respectively.
Also, $\eta$ and $\xi$ are the shear viscosity and bulk viscosity,
respectively.

To solve the set of equations~\eqref{lhe}, Ernst \emph{et al.}
\cite{ernst1971} use the Fourier transform with respect to space.
In this paper, we use the Fourier series of the hydrodynamic variables
with respect to space~\cite{keyes1975,erpenbeck1982}, given by
\begin{equation}
 \begin{split}
  \delta n \left(\boldsymbol{r},t \right) & = \frac{1}{V} \sum_{\boldsymbol{k}}
  e^{i \boldsymbol{q} \cdot \boldsymbol{r}} n_{\boldsymbol{q}}  \left(t \right),  \\
  n_{\boldsymbol{q}}  \left(t \right) & = \int_V d\boldsymbol{r} \,,
  e^{- i \boldsymbol{q} \cdot \boldsymbol{r}} \delta n \left(\boldsymbol{r},t \right),
 \end{split}
\end{equation}
where $\delta n \equiv n (\boldsymbol{r},t ) - n$ is the
deviation from the equilibrium value, and 
the system is a rectangular solid
$\{ L_x,L_y,L_z \}$, ${q}_i \equiv 2\pi k_i/L_i$
($i = x,y,z$), with integer $k_i$.
Similarly, we introduce the Fourier components $T_{\boldsymbol{q}} (t)$
and $\boldsymbol{u}_{\boldsymbol{q}} (t)$, which are the components
of $\delta T \left(\boldsymbol{r},t \right) = T \left(\boldsymbol{r},t \right) - T$
and $\boldsymbol{u} \left(\boldsymbol{r},t \right)$, respectively.

In $d$-dimensional Fourier space, Eq.~\eqref{lhe} is converted into 
\begin{equation}
 \label{lhek}
 \begin{split}
  \frac{\partial n_{\boldsymbol{q}}}{{\partial t}} = &
  - i n \boldsymbol{q} \cdot {\boldsymbol{u}_{\boldsymbol{q}}},\\
  \frac{\partial \boldsymbol{u}_{\boldsymbol{q}}}{{\partial t}} = &
   - \nu \boldsymbol{q}^2 \boldsymbol{u}_{\boldsymbol{q}}
  - \left(D_l-\nu \right)\boldsymbol{q}  \left(\boldsymbol{q} \cdot \boldsymbol{u}_{\boldsymbol{q}} \right) \\
  & \,\,\,\,\, - \frac{i \boldsymbol{q} c_0^2}{\gamma}
   \left(\frac{n_{\boldsymbol{q}}}{n} + \alpha_p T_{\boldsymbol{q}} \right) ,\\
  \frac{\partial T_{\boldsymbol{q}}}{{\partial t}} = &
  - \frac{\gamma -1}{\alpha_p} i \boldsymbol{q} \cdot \boldsymbol{u}_{\boldsymbol{q}}
  - \gamma D_T q^2 T_{\boldsymbol{q}}.
 \end{split}
\end{equation}
These equations can be reduced to a diagonalizable 
$(d+2)$-dimensional matrix equation.
Here, we introduce the ``hydrodynamic modes'', defined as the
eigenfunctions of the form $\exp (- \omega_q^j t)$.
The quantities $\omega_q^j$ in the small $q \,\, (\equiv |\boldsymbol{q}| )$ limit
are calculated as the $d-1$ viscous modes
[$\omega_q^j = \nu q^2 \,\,\, (j = 1, \cdots , d-1)$],
the two sound modes
($\omega_q^{\pm} = \pm i c_0 q +\frac{1}{2} \Gamma_s$), 
with
\begin{equation}
 \label{def-gammas}
 \Gamma_s \equiv D_l +  \left(\gamma-1 \right) D_T,
\end{equation}
and the heat mode ($\omega_q^H = D_T q^2 $).
The Fourier series of the hydrodynamic variables can be expressed
as linear combinations of the eigenvectors of the matrix equation.

After setting the initial values used by Ernst \emph{et al.}~\cite{ernst1971},
$C_\lambda (t)$ can be calculated.
When the system is sufficiently large,
the most dominant part of the long-time tail of $C_\lambda (t)$ is the
viscous-heat (VH) term, given by 
\begin{equation}
 \label{viscous-heat}
  C_\lambda^{VH}  \left(t \right) \sim \frac{1}{V} \sum_{\boldsymbol{k}}
  \left(1-\hat{q}_x^2 \right)
  \exp\left[- \left(\nu+D_T \right) q^2 t\right],
\end{equation}
with $\hat{q}_i \equiv q_i/q \,\, \left(i = x,y,z \right)$
and the sound-sound (SS) term,
\begin{equation}
 \label{sound-sound}
 C_\lambda^{SS} (t) \sim \frac{1}{V} \sum_{\boldsymbol{k}} \hat{q}_x^2
  \exp \left[- \Gamma_s q^2 t\right].
\end{equation}
Note that $C_\lambda (t)$ decays as $t^{-d/2}$~\cite{ernst1971,ernst1976}.

In a Q1D system, however, the SS term is dominant,as seen from the following. 
First, let us assume that $L_y \simeq L_z$ and
that $L_x$ is sufficiently larger than $L_y$ and $L_z$. 
The exponential term in the SS term in Eq.~\eqref{sound-sound} is
\begin{equation}
 \exp \left[- 4 \pi^2 \Gamma_s
       \left(\frac{k_x^2}{L_x^2} + \frac{k_y^2}{L_y^2}
	+ \frac{k_z^2}{L_z^2}  \right) t\right].
\end{equation}
When time is sufficiently large, \emph{i.e.}, $t \sim t_c \sim L/c_0$ with
$L = \sqrt{L_x^2+L_y^2+L_z^2} \simeq L_x$,
it is found that the $\{k_y, k_z\} = \{0,0 \}$ term is dominant
in Eq.~\eqref{sound-sound}.
Because we consider Q1D systems, we assume that the relation
given by 
\begin{equation}
 \label{nc0}
  \frac{\Gamma_s t}{L_y^2} \sim \frac{\Gamma_s}{c_0 L_y^2} L_x \gg 1
\end{equation}
is satisfied.
Therefore Eq.~\eqref{sound-sound} can be rewritten as
\begin{equation}
 C_\lambda^{SS}  \left(t \right) \sim \frac{1}{L_y L_z} \frac{1}{L_x}
 \sum_{k_x} \exp \left[- 4 \pi^2 \Gamma_s \frac{k_x^2}{L_x^2} t\right].
 \label{sound-sound-c}
\end{equation}
If $\Gamma_s t/L_x^2 \sim \Gamma_s / (c_0 L_x)$
is sufficiently small, 
this summation can be reduced to the integral form
\begin{equation}
 \label{sound-sound-t}
  \begin{split}
   C_\lambda^{SS} (t) & \sim \frac{1}{L_y L_z} \int_{0}^{\infty} dx
   \,\, \exp \left[- 4 \pi^2 \Gamma_s x^2 t\right] , \\
   & = \frac{1}{4 \pi L_y L_z} \sqrt{\frac{\pi}{\Gamma_s t}},
  \end{split}
\end{equation}
which implies that $C_\lambda^{SS} (t)$ decays as $t^{-1/2}$.

By contrast,
the $\{k_y, k_z\} = \{0,0 \}$ term in the VH term of Eq.~\eqref{viscous-heat}
cannot be dominant. 
Instead, the $\{k_y,k_z \} = \{ \pm 1,0 \}$ and $\{0, \pm 1 \}$ terms
are dominant, and 
Eq.~\eqref{viscous-heat} can be written in the integral form 
\begin{equation}
 \label{viscous-heat-t}
 \begin{split}
  C_\lambda^{VH}
  \left(t \right) & \sim \frac{1}{L_y L_z} \int_{0}^{\infty} dx
  \,\, \frac{1}{1+L_y^2 x^2}
  \exp \left[- 4 \pi^2  (\nu+D_T )
  \frac{1 + L_y^2 x^2}{L_y^2} t \right], \\
  & <
  \frac{1}{L_y L_z} \exp \left[ - 4 \pi^2  \frac{\nu+D_T }{L_y^2} t \right]
  \int_{0}^{\infty} dx \,\, \exp \left[- 4 \pi^2
  \left(\nu+D_T \right) x^2 t\right], \\
  & =
  \frac{1}{4 \pi L_y L_z} \sqrt{\frac{\pi}{(\nu+D_T)t}}
  e^{- 4 \pi^2  \frac{\nu+D_T }{L_y^2} t} ,
 \end{split}
\end{equation}
which reveals that $C_\lambda^{VH} (t)$ decays exponentially with time.
Thus, $C_\lambda^{VH} (t)$ decays much faster than $C_\lambda^{SS} (t)$.
From Eq.~\eqref{tc}, if is seen that the heat conductivity can be calculated as
\begin{equation}
 \lambda \simeq \beta T^{-1} \int^{t_c}_0 dt
  \,\, \left(C_\lambda^{SS} \left(t \right)
	+ C_\lambda^{VH}  \left(t \right) \right)
  \simeq \beta T^{-1} \int^{t_c}_0 dt C_\lambda^{SS} (t).
  \label{test1}
\end{equation}

From Eqs.~\eqref{sound-sound-t} and~\eqref{test1},
we need to estimate the system size dependence of $\Gamma_s$
in order to calculate $\lambda$.
Therefore, we need to calculate the
long-time tail of the shear viscosity in Q1D systems.
The time correlation function of
the $xy$ component of the shear stress and the corresponding viscosity
are calculated as
\begin{eqnarray}
 \label{eta}
  \eta^{xy} &=& \frac{\beta}{V} \int^{t_c}_0 dt \,\, C_\eta^{xy} (t) , \\
 C_\eta^{xy} (t) &\equiv& 
  \meanvalue{J_\eta^{xy} (t) J_\eta^{xy} (0)} , \\
 J_\eta^{xy} (t) &\equiv&
  \sum_{i}
  \left(m v_{i x} (t) v_{i y} (t) - \frac{1}{2}
  \sum_{j\neq i} r_{ij,x} \frac{\partial \Phi(\boldsymbol{r}_{ij})}{\partial r_{ij,y}}
  \right) , \\
 J_\eta^{K,xy} &=&
  \sum_{i} m v_{i x} (t) v_{i y} (t).
\end{eqnarray}
We consider only the kinetic parts, $J_\eta^{K,xy}$, of $J_\eta^{xy}$
and the corresponding correlation function.
From the similar calculation of $C_\lambda (t)$, if is found that the dominant term
of $C_\eta (t)$ for large $L_x$ is the viscous-viscous (VV) term from
the $yz$ component of the shear stress, given by 
\begin{equation}
 \begin{split}
 {C_\eta^{yz}}^{VV} (t) \sim 
  \sum_{\boldsymbol{k}}
  [\hat{q}_x^2 + 2 \hat{q}_y^2\hat{q}_z^2]
  \exp [- 2 \nu \boldsymbol{q}^2 t] \sim 
  L_x \sqrt{\frac{1}{\nu t}}.
 \end{split}
\end{equation}
Because the kinetic viscosity $\nu$
is proportional to $\eta$ in Eq.~\eqref{def-variable},
we derive $\eta$ from Eq.~\eqref{eta} as
\begin{equation}
 \label{asv}
  \eta \sim \left(\frac{L_x}{L_y^2 L_z^2} \right)^{1/3}.
\end{equation}
We assume that the bulk viscosity $\xi$ satisfies the relation
\begin{equation}
 \eta + \xi \sim \left(\frac{L_x}{L_y^2 L_z^2} \right)^{1/3},
\end{equation}
which is realized in most normal fluids.
Therefore, $\Gamma_s$ can be expressed as
\begin{equation}
 \label{gammas}
  \Gamma_s \sim C_1 \lambda + C_2 \left(\frac{L_x}{L_y^2 L_z^2}\right)^{1/3},
\end{equation}
where $C_1$ and $C_2$ are constants that
are independent of time and the system size.
Thus, from Eqs.~\eqref{sound-sound-t}, \eqref{test1} and
\eqref{gammas}, we can derive the following self-consistent equation for 
$\lambda$:
\begin{equation}
 C_1 \lambda^3 + C_2 \left(\frac{L_x}{L_y^2 L_z^2} \right)^{1/3} \lambda^2
  \sim \frac{L_x}{L_y^2 L_z^2}.
\end{equation}
This equation indicates that the system size dependence of $\lambda$ is
given by
\begin{equation}
 \label{ahc}
 \lambda \sim \left(\frac{L_x}{L_y^2 L_z^2}\right)^{1/3}.
\end{equation}
The necessary condition for Eq.~\eqref{ahc} to be valid is derived from
Eq.~\eqref{nc0} as
\begin{equation}
 \label{nc1}
  \phi \gg 1.
\end{equation}

The heat conductivity and shear viscosity
in Q2D systems can be derived correspondingly as
\begin{equation}
 \lambda \sim \sqrt{\frac{\ln (L_x^2+L_y^2)}{L_z}},
\end{equation}
where $L_z \ll L_x, L_y$. 
The necessary condition for this anomalous behavior is derived as
\begin{equation}
 \frac{L_x^2 + L_y^2}{n L_z^5} \ln \frac{\sqrt{L_x^2+L_y^2}}{t_0 c_0} \gg 1.
\end{equation}

\section{MD simulation} \label{simulation}

In order to check the validity of Eq.~\eqref{ahc},
we have performed 3D MD simulations with hard spheres in Q1D systems
\cite{rapaport1980,marin1993,marin1995,isobe1999}.
It is noted that 
a logarithmic system size dependence of the heat conductivity
in 2D systems has been confirmed
with MD simulations with hard disks \cite{murakami2003}.
It should also be noted, however, that it is difficult to distinguish
logarithmic behavior from power law behavior in 2D simulations.

\begin{figure}[t]
 \centering
 \includegraphics[width=80mm]{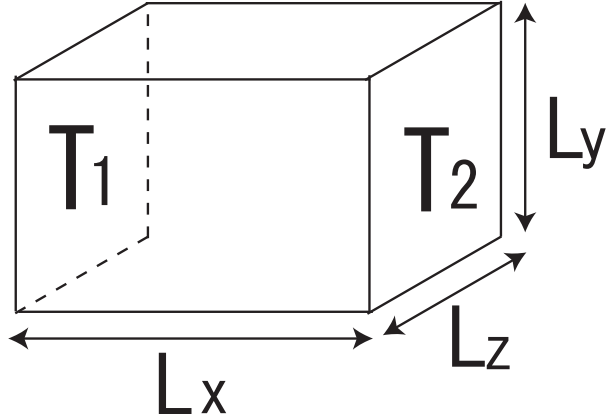}
 \caption{The system studied in the simulation.
 The walls perpendicular to the $y$ and $z$ directions
 are perfectly reflecting
 and those perpendicular to the $x$ direction are perfectly thermalizing.
 These `thermal' walls are at $T_1$ and $T_2$, respectively.}
 \label{system}
\end{figure}

In our simulations, the system is confined in a box of size
${L_x \times L_y \times L_z}$, as shown in Fig.~\ref{system}.
The walls perpendicular to the $y$ and $z$ directions
are perfectly reflecting walls and the walls vertical to
the $x$ direction are perfectly thermalizing walls at temperatures
$T_1$ and $T_2$, respectively.
At the `thermal' wall of temperature $T_w$,
a particle is reflected with a new velocity $\boldsymbol{v}$ at
random.
The probability distribution function for $\boldsymbol{v}$ is given by
\begin{equation}
 \psi  \left(\boldsymbol{v} \right)
  = \frac{2 \pi |v_x|}{ \left(2 \pi k_B T_w \right)^{2}}
  \exp\left[-\frac{\boldsymbol{v}^2}{2 k_B T_w}\right].
\end{equation}

The MD data were taken for the systems with
$\tilde{L}_y = \tilde{L}_z = 0.1 - 3.2$ and
$\tilde{L}_x/\tilde{L}_y = 2 - 16384$, where
the unit of length is the diameter of a particle and
$\tilde{L}_i \equiv L_i - 1$ is the range within which the center of the
particle can move. 
For the system with $\tilde{L}_y = 3.2$, data were
taken only for $\tilde{L}_x/\tilde{L}_y = 2 - 4096$.
The packing fraction was fixed approximately to $0.055$.
The ratio of the temperature difference between two thermal walls 
was $T_1/T_2 = 2$.

As the initial state,
the molecules were arranged so as 
to realize a constant pressure under the linear
profile of $T$; i.e. $nT$ is initially uniform in the system.
The initial velocities of the particles were chosen from the Maxwellian 
distribution with $T$ varying linearly from $T_1$ to $T_2$.
To obtain a stationary initial state,
we simulated $6 \times 10^5$ collisions per particle which
started from the initial state.
We carried out the simulation until $4 \times 10^5$ collisions
per particle were realized from the stationary initial state.
The heat flux $J$ was calculated using Eq.~\eqref{jj}.
We found that in the simulation, the difference between 
$J$ and $J_K$ is less than $0.1 \%$.
The heat conductivity is calculated as
$\lambda = - J \hat{L}_x / \triangle T$, where $\triangle T$ is the
difference between the temperatures 
near the thermal walls at $T_1$ and $T_2$, respectively, and $\hat{L}_x$ is
the distance between the points at which the temperatures are given.

\begin{figure}[t]
 \centering
 \includegraphics[width=80mm]{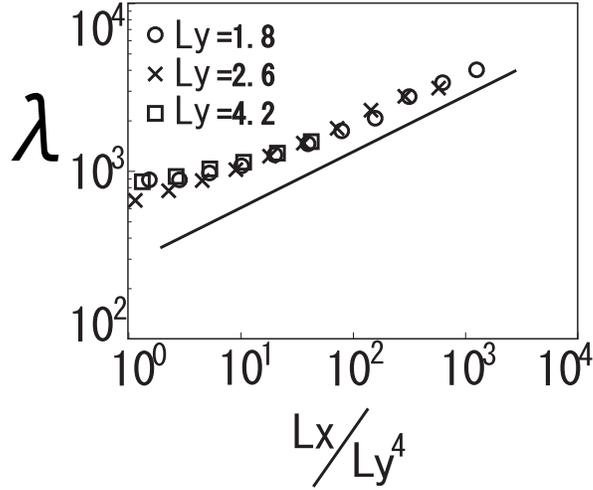}
 \caption{The size dependence of the heat conductivity in the Q1D system.
 The $x$-axis is $L_x/L_y^4$,
 with the unit of length being the diameter of the hard spheres.
 The circles,
 crosses and open squares indicate the heat conductivities, where $L_y$ is
 $1.8, 2.6$ and $4.2$, respectively. The solid line represents
 a function proportional to $x^{1/3}$.}
 \label{result}
\end{figure}

\begin{figure}[t]
 \centering
 \includegraphics[width=120mm]{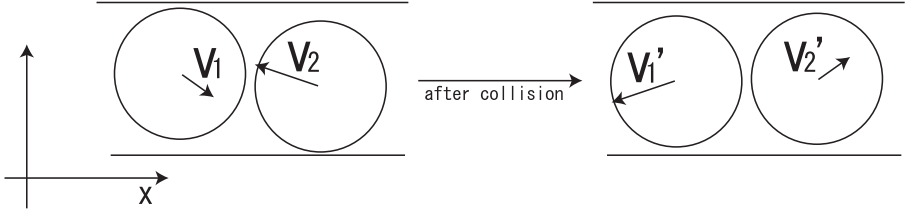}
 \caption{A collision
 of particle $1$ and particle $2$ for small $L_y$.
 Here, $\boldsymbol{v}_1$ and $\boldsymbol{v}_2$ are the
 velocities of particles $1$ and $2$ before the collision,
 respectively, and
 $\boldsymbol{v}'_1$ and $\boldsymbol{v}'_2$ are the velocities
 of particles $1$ and $2$
 after the collision, respectively.}
 \label{collision}
\end{figure}

The result of the simulation is shown in Fig.~\ref{result}.
There, it is seen that the heat conductivity seems to obey Eq.~\eqref{ahc}
for $L_y \geq 1.8$, while such behavior is not observed for smaller $L_y$.
We believe that this difference is caused by
the limitation on the collision angle in the case that $\tilde{L}_y$ is
much smaller than the diameter.
For small $\tilde{L}_y$, as in Fig.~\ref{collision},
most of the collisions of the particles take place with
$v'_{1x} \simeq v_{2x}$ and $v'_{2x} \simeq v_{1x}$, and
the changes of the velocities in the other directions
are small in each collision.
Thus, most of the collisions cause
only exchanges of the velocities of the particles in
the $x$-direction, and the heat transfer behaves like that in the case
of ballistic transport.
In this situation, the linearized hydrodynamic equations~\eqref{lhe}
for small $L_y$ cannot be used.
Contrastingly, for the system with $L_y = 1.8$ and $L_x/(L_y^2 L_z^2) > 10.0$,
the most probable value of the exponent
$\alpha$ with $L^{\alpha}$ is $0.30 \pm 0.01$, which is close 
to $1/3$.

Although the results of our theory and simulation are consistent,
we need to simulate larger systems to confirm the relation~\eqref{ahc}.
There is an inconsistency in the analytical and numerical treatments.
Indeed, because the linearized hydrodynamic equations~\eqref{lhe} were
adopted in the case that the system size is larger than the
mean free path, but this conditions is not satisfied in our simulation,
where the widths $L_y$ and $L_z$ are smaller than the mean free path
by an amount on the order of $7$ diameters.

\section{Discussion}

We should note that Shimada \emph{et al.} \cite{shimada2000} and Ogushi
\emph{et al.} \cite{ogushi2005}
also calculated the heat conductivity
using 3D MD simulations in Q1D systems consisting of hard spheres and Lennard-Jones
molecules, respectively.
Their simulations suggest that the heat conductivity
does not diverge in the limit $L_x \to \infty$.
However, the system sizes in their simulations do not satisfy
the necessary condition $\phi \gg 1$.
In their simulations,
$\phi$ is $0.03 - 0.3$
with $L_y = 20.0$ and $0.3 - 2$ with $L_y = 4.0$, respectively,
where the unit of the length is the particle diameter.
By contrast, our simulation is more extensive 
in the system $\phi$ is $1.8 - 2 \times 10^{4}$ for $L_y = 1.8$.
This difference might be the cause of the qualitative difference in the
behavior of the heat conductivity.

\begin{figure}[t]
 \centering
 \includegraphics[width=80mm]{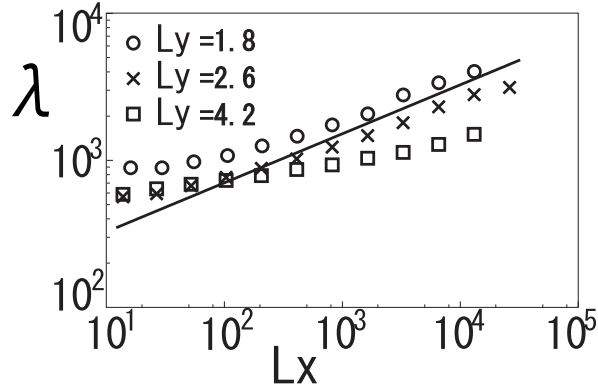}
 \caption{The size dependence of the heat conductivity in the Q1D system.
 The $x$-axis is $L_x$,
 with the unit of length being the diameter of the hard spheres.
 The circles,
 crosses and open squares indicate the heat conductivities, where $L_y$ is
 $1.8, 2.6$ and $4.2$, respectively. The solid line represents
 a function proportional to $x^{1/3}$.}
 \label{result2}
\end{figure}

In numerical simulations of SWNT, which can be regarded as a Q1D system,
$\lambda \sim L^{1/3}$ has been observed \cite{murayama2002,zhang2005}.
In Murayama's paper \cite{murayama2002}, it is reported that
the heat conductivity behaves as $\lambda \sim L_l^{0.32}$, with the tube length $L_l$
for $(5,5)$ SWNT. However,
for a wider tube diameter, $(10,10)$ SWNT, the increase of $\lambda$
is suppressed.
It might be possible to understand this result in terms of the necessary condition
for the anomalous behavior of the heat conductivity, $\phi \gg 1$.
This behavior is also consistent with that found in our simulation.
As we can see in Fig.~\ref{result2},
anomalous behavior of the heat conductivity, $\alpha \simeq 1/3$,
with $\lambda \sim L_x^{\alpha}$, is observed for $L_x > 10^3$, 
with $L_y = 1.8$ where
$\phi$ is larger than the critical value,
$\phi_c = 7.1 \times 10^2$.
By contrast, for a wider tube diameter, $L_y = 4.2$, the increase of
$\lambda$ is suppressed.
From the necessary condition \eqref{nc1}, the anomalous
behavior $\alpha \simeq 1/3$ may be observed for $L_x > 8.4 \times 10^3$
for the wider tube diameter $L_y = 4.2$,
for which $\phi > \phi_c$ is satisfied.
Similarly, for $(10,10)$ SWNT, anomalous behavior of the heat
conductivity might be observed for a longer tube,
because $\phi \gg 1$ is satisfied in that case.

Recently, Shiba \emph{et al.} \cite{shiba2006} observed
that the time correlation function of the heat flux
is proportional to $t^{-0.98 \pm 0.25}$, and the heat conductivity exhibits logarithmic
divergence with the system size in three-dimensional FPU-$\beta$ lattice systems.
They also measured the size dependence of the heat conductivity in Q1D
FPU-$\beta$ lattices.
Figure $4$ in their paper suggests that when the system is sufficiently
close to a Q1D system, the heat conductivity is proportional to $L^{1/3}$.

In conclusion, we derived relation according to which 
the heat conductivity is proportional to $(L_x/L_y^2 L_z^2)^{1/3}$ in Q1D systems 
and $\sqrt{(\ln {(L_x^2+L_z^2)})/L_y}$ in Q2D systems, and
the necessary conditions $L_x^4/(n^2 L_y^5 L_z^5) \gg 1$ and 
\hspace{27mm} \,
$((L_x^2 + L_y^2) / (n L_z^5)) \ln
[\sqrt{L_x^2 + L_y^2}/(t_0 c_0)] \gg 1$, respectively, 
for the anomalous behavior of the heat conductivity from linearized
hydrodynamic equations.
This behavior in Q1D systems has been confirmed through comparison with
MD simulations.

\section*{Acknowledgements}
The author would like to express his sincere gratitude to
Prof. H. Hayakawa for his valuable comments and 
critical reading of the manuscript.
The author also thanks
Prof. S. Takesue and Prof. S. Sasa
for valuable discussions and comments.
This work is partially
supported by a Grant-in-Aid from the Japan Space Forum
and the Ministry of Education,
Culture, Sports, Science and Technology (MEXT),
Japan (Grant No.18540371) and a Grant-in-Aid from the 21st century
COE ``Center for Diversity and Universality in Physics'' from MEXT, Japan.
The numerical computations reported in this work were carried out at the 
Yukawa Institute Computer Facility.

%


\begin{thebibliography}{99}
  
\bibitem{lepri2003}
	S.~Lepri, R.~Livi and A.~Politi, \PRP{377,2003,1}.
 \bibitem{shimada2000}
	T.~Shimada, T.~Murakami, S.~Yukawa, K.~Saito and N.~Ito,
	\JPSJ{69,2000,3150}.
 \bibitem{ogushi2005}
	F.~Ogushi, S, Yukawa and N.~Ito,
	\JPSJ{74,2005,827}.
 \bibitem{narayan2002}
	O.~Narayan and O.~Ramaswamy,
	\PRL{89,2002,200601}.
 \bibitem{mai2006}
	T.~Mai and O.~Narayan,
	\PRE{73,2006,061202}.
 \bibitem{wang2004}
	J.-S.~Wang and B.~Li,
	\PRL{92,2004,074302}.
 \bibitem{grassberger2002}
	P.~Grassberger, W.~Nadler and L.~Yang,
	\PRL{89,2002,180601}.
 \bibitem{cipriano2005}
	P.~Cipriano, S.~Denisov and A.~Politi,
	\PRL{94,2005,244301}.
 \bibitem{murayama2002}
	S.~Murayama, \JL{Physica B,323,2002,193}.
 \bibitem{zhang2005}
	G.~Zhang and B.~Li,
	\JL{J.~Chem.~Phys.,123,2005,014705}.
 \bibitem{alder1970}
	B.~J.~Alder and T.~E.~Wainwright,
	\PRA{1,1970,18}.
 \bibitem{pomeau1975}
	Y.~Pomeau and P.~Resibois,
	\PRP{19,1975,64}.
 \bibitem{ernst1971}
	M.~H.~Ernst, E.~H.~Hauge and J.~M.~J.~van Leeuwen,
	\PRA{4,1971,2055}.
 \bibitem{ernst1976}
	M.~H.~Ernst, E.~H.~Hauge and J.~M.~J.~van Leeuwen,
	\JL{J.~Stat.~Phys.,15,1976,7}.
 \bibitem{kubo1957}
	R.~Kubo,
	\JPSJ{12,1957,570}.
 \bibitem{zwanzig1965}
	B.~Zwanzig,
	\JL{Annu.~Rev.~Phys.~Chem.,16,1965,67}.
 \bibitem{keyes1975}
	T.~Keyes and B.~Ladanyi,
	\JL{J.~Chem.~Phys.,62,1975,4787}.
 \bibitem{erpenbeck1982}
 	 J.~J.~Erpenbeck and W.~W.~Wood,
	 \PRA{26,1982,1648}.
 \bibitem{rapaport1980}
	 D.~C.~Rapaport,
	 \JL{J.~Comput.~Phys.,34,1980,184}.
 \bibitem{marin1993}
	 M.~Mar\'{i}n , D.~Risso and P.~Cordero,
	 \JL{J.~Comput.~Phys.,109,1993,306}.
 \bibitem{marin1995}
	 M.~Mar\'{i}n and P.~Cordero,
	 \JL{Comput.~Phys.~Commu.,92,1995,214}.
 \bibitem{isobe1999}
	M.~Isobe,
	\JL{Int.~J.~Mod.~Phys. C, 10,1999,1281}.
 \bibitem{murakami2003}
	T.~Murakami, T.~Shimada, S.~Yukawa and N.~Ito,
	\JPSJ{72,2003,1049}.
 \bibitem{shiba2006}
	H.~Shiba, S.~Yukawa and N.~Ito,
	\JPSJ{75,2006,103001}.
\end{thebibliography}
\end{document}